\documentclass[11pt]{article}
\usepackage{moriond,epsfig}

\def\PLB{{\em Phys. Lett.}  B}
\def\PRL{\em Phys. Rev. Lett.}

\def\PRC{{\em Phys. Rev.} C}

\def\be{\begin{equation}}
\def\ee{\end{equation}}
\def\bea{\begin{eqnarray}}
\def\eea{\end{eqnarray}}

\begin{document}
\vspace*{4cm}
\title{Measurements of J/$\psi$ yields at forward and mid-rapidity in d+Au,
Cu+Cu and Au+Au collisions at $\sqrt{s_{NN}}$~=~200~GeV by PHENIX at RHIC}

\author{ VN Tram }

\address{LLR - Ecole Polytechnique\\
Route de Saclay 91128 Palaiseau, France}

\maketitle\abstracts{
The J/$\psi$ yields in heavy ion collisions are expected to be a promising probe of deconfined matter, since theoretical models predict that the J/$\psi$ production could be strongly suppressed due to color screening effect in a Quark Gluon Plasma. In addition, one of the interesting predictions that has emerged recently is the J/$\psi$ enhancement at RHIC energy via competing mechanisms such as recombination. The PHENIX collaboration has measured J/$\psi$ production via its decay into lepton pairs in Au+Au and Cu+Cu collisions at $\sqrt{s_{NN}}$ = 200 GeV. The nuclear modification factor $R_{\rm AA}$ for the ${\rm J}/\psi$ is obtained by comparing Au+Au or Cu+Cu collisions to p+p collisions. The $R_{\rm AA}$ dependences on centrality, transverse momentum and rapidity are presented both at forward rapidity ($1.2<|y|<2.2$) using ${\rm J}/\psi\rightarrow\mu^+\mu^-$ and at mid rapidity ($|y|<0.35$) using ${\rm J}/\psi\rightarrow e^+e^-$. These preliminary results are compared to cold nuclear matter expectations derived from PHENIX d+Au measurements, to results obtained by NA50 experiment and to various theoretical models.}

\section{J/$\psi$ production in heavy ion collisions by the PHENIX experiment}
In heavy ion collisions, the J/$\psi$ production is used as a probe of the produced medium. In the case of Quark Gluon Plasma formation, the J/$\psi$ yield is expected to be suppressed. The NA50 Experiment at the SPS has already measured a suppression in Pb+Pb collisions~\cite{psi_result_na50}: this suppression is higher than the suppression due to the normal nuclear absorption and is thus called anomalous. However, models without QGP formation also describe the observed suppression pattern and amplitude~\cite{comovers}. At RHIC, the PHENIX experiment aims to observe this effect at a 10 times higher collision energy. The PHENIX experiment is able to measure J/$\psi$ in 2 different rapidity regions with magnetic spectrometers: at mid-rapidity in the dielectron decay channel ($|y|<0.35$) and at forward rapidity in the dimuon decay channel ($1.2<|y|<2.2$). Electrons are identified in the central arms by their {\v C}erenkov rings and by matching the momentum of charged particles reconstructed in drift chambers with the energy deposited in an electromagnetic calorimeter. Muons are selected by an absorber and identified by the penetration depth they reach in a succession of proportional counters staggered with steel walls.\\
At RHIC energy, J/$\psi$'s are mainly produced by gluon fusion. In addition to directly produced J/$\psi$'s, measurements also include decays from higher mass  $\chi_c$ and $\psi'$ resonances. The J/$\psi$ production in any A+B ions system is expected to be the production measured in p+p system scaled by the number of binary nucleon-nucleon collisions. Thus, results will be shown as the nuclear modification factor $R_{AB}$, defined as the production measured in A+B collisions over the production in p+p times the mean number of binary collisions corresponding to the centrality bin used in A+B. However, heavy ion collisions introduce cold nuclear effects such as gluon shadowing (modification of the gluon structure function of nucleons in nuclei), transverse momentum broadening (the so-called Cronin effect) in the initial state (before the $c\bar{c}$ formation) and nuclear absorption in the final state. Measuring the J/$\psi$ yield in d+Au collisions enables to quantify these effects. At RHIC energies, results from the d+Au data set~\cite{ppg038} indicate a weak shadowing and anti-shadowing, which can be described by EKS98 parametrization~\cite{vogt_dAu}. The results also point out that the nuclear absorption is weak ($\sigma_{abs}$ = 1 to 3~mb) as compared to the one measured at the SPS ($\sigma_{abs}$ = 4.18~mb). In this presentation, $J/\psi$ results~\cite{hugo} obtained in Cu+Cu (3.06 $nb^{-1}$) and Au+Au (241 $\mu b^{-1}$) collisions are reported and discussed. Data samples from d+Au and p+p collisions shown in this paper, have been published by PHENIX~\cite{ppg038}.

\section{Results and first interpretations}
\begin{figure}[htbp!]
\vspace{-0.45cm}
\begin{minipage}{7.6cm}
\begin{center}\includegraphics[width=6.5cm,keepaspectratio]{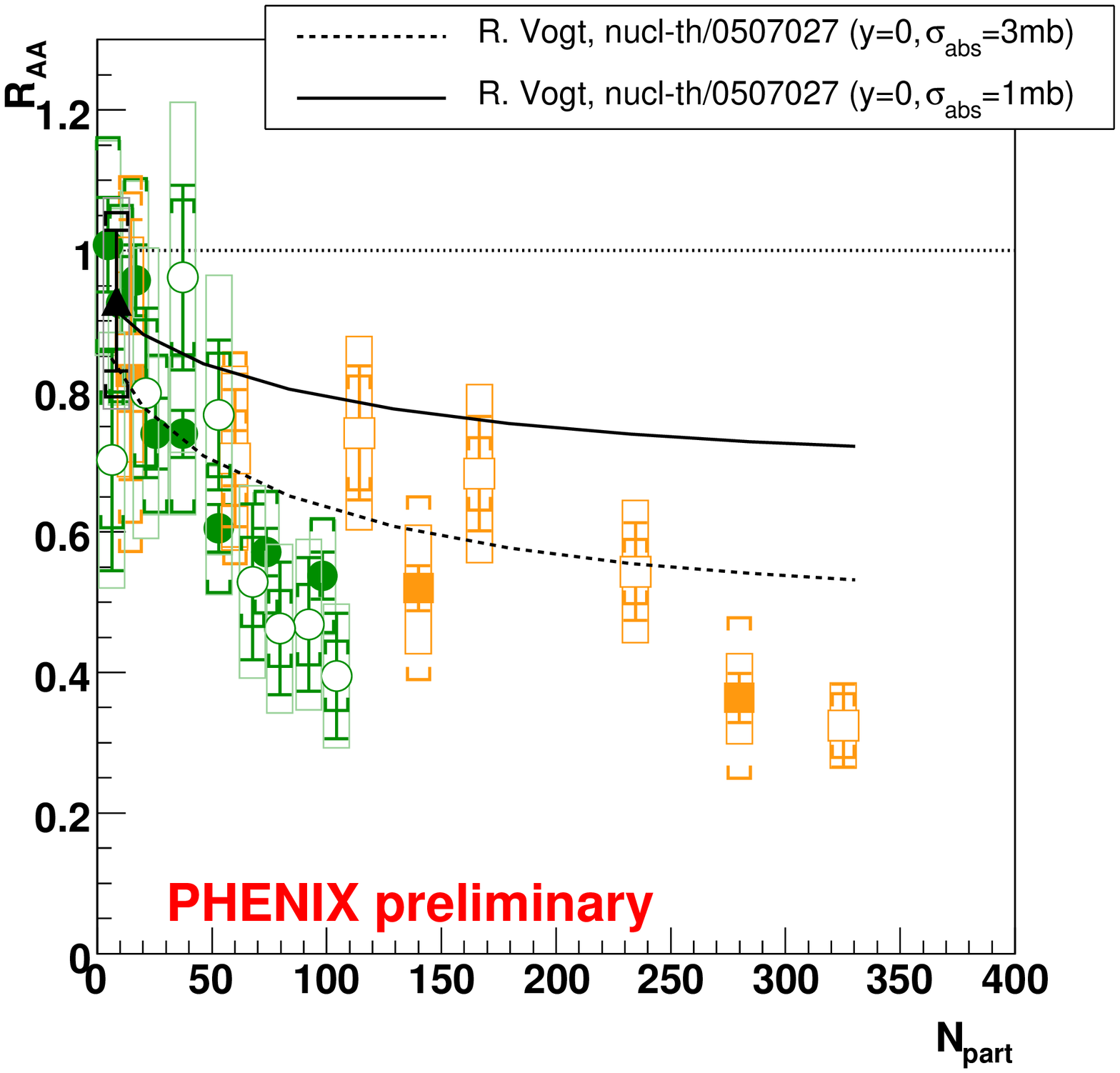}
\end{center}
\end{minipage}
\begin{minipage}{7.6cm}
\vspace{-0.5cm}
\begin{center}
\includegraphics[width=6.7cm,keepaspectratio]{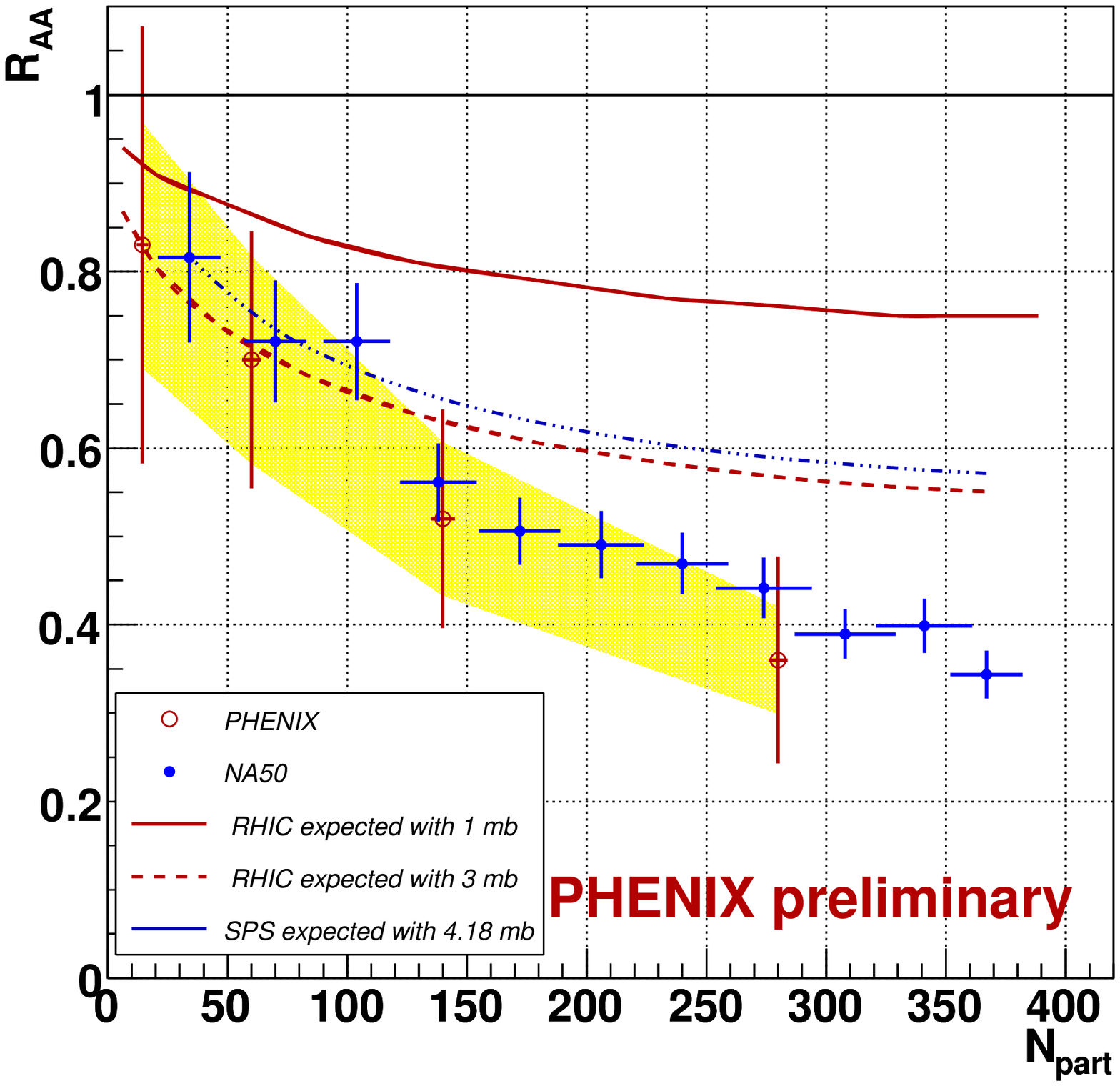}
\end{center}
\end{minipage}
\vspace{-0.4cm}
\caption{\label{raa_all}{Left: Nuclear modification factor $R_{AA}$ as a function of $N_{part}$, measured in d+Au (black triangle)~{\protect\cite{ppg038}}, Cu+Cu (green circles) and Au+Au (gold squares) collisions at foward rapidity $1.2<|y|<2.2$ (filled markers) and at mid-rapidity $|y|<0.35$ (open markers)~{\protect\cite{hugo}}. Black curves correspond to the expected cold nuclear effects in Au+Au collisions~{\protect\cite{vogt}} using EKS98 parametrization and a nuclear absorption cross section $\sigma_{abs}$ = 1 mb (solid line) and 3 mb (dashed line). Right: J/$\psi$ nuclear modification factor as a function of centrality measured by PHENIX (empty circles) and by NA50 (dot)~{\protect\cite{psi_result_na50}}}. The blue dot-dashed line corresponds to the J/$\psi$ expected production in NA50($\sigma_{abs}$ = 4.18 mb). The red dashed (solid) line corresponds to the prediction for PHENIX (shadowing (EKS98) and nuclear absorption $\sigma_{abs}$ = 3 (1) mb).}
\vspace{-0.4cm}
\end{figure}

The nuclear modification factor $R_{AB}$ is presented in figure~\ref{raa_all} as a function of centrality (the number of nucleon participants $N_{part}$) for d+Au, Cu+Cu and Au+Au collisions. In the most central Au+Au collisions, the J/$\psi$ yield is suppressed by a factor of 3 as compared to the binary scaling. This suppression is larger the cold nuclear effect in Au+Au collisions~\cite{vogt}, suggesting that other mechanisms are involved.\\

The comparison between NA50 and PHENIX results is an intricate problem. The collision energy is 10 times higher at RHIC, leading to a higher temperature which could affect the J/$\psi$ production differently. Furthermore, the cold nuclear effects may be different: due to a different incident energy the nucleus overlap time scale is different between SPS and RHIC, this can modify the J/$\psi$ nuclear absorption. The different Bjorken $x$ domains also lead to different shadowing or anti-shadowing effects. The right figure~\ref{raa_all} presents the nuclear modification factor $R_{AA}$ as a function of the number of participants for NA50~\footnote{Normalized with p+p collisions studied by NA51~\cite{na51}.} and PHENIX. The suppression amplitudes are very similar which is surprising. However, the cold nuclear effects are expected to occur differently at the SPS and at RHIC. When using a cross section of 3 mb at RHIC to describe the nuclear absorption, the measured over expected ratio amplitude becomes quite similar between the 2 experiments, suggesting a possible scenario in which the temperature at RHIC is not high enough to reach direct J/$\psi$ suppression~\footnote{It is not excluded that direct J/$\psi$ suppression temperature could be reached at RHIC: more statistics would allow to use smaller centrality bins and thus to probe a possible direct J/$\psi$ suppression at higher temperature.}. In this case, the measured suppression would be due to the sequential suppression of higher mass resonances ($\chi_c$ and $\psi'$  decaying into J/$\psi$). For a more detailed comparison, a better understanding of the cold nuclear effects at RHIC is mandatory, especially one needs to better constrain the normal nuclear absorption cross section. \\

\begin{figure}[htbp!]
\vspace{-0.6cm}
\begin{minipage}{8.0cm}
\begin{center}\includegraphics[width=6.5cm,keepaspectratio]{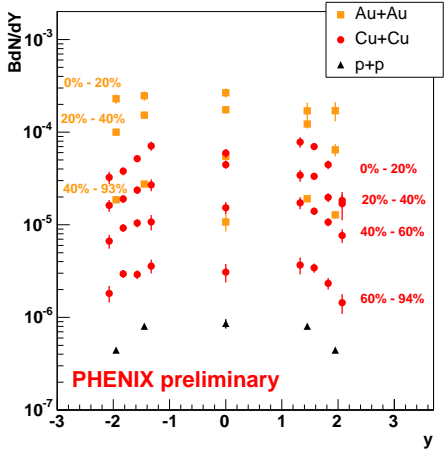}
\end{center}
\end{minipage}
\begin{minipage}{7.7cm}
\begin{center}
\includegraphics[width=6.5cm,keepaspectratio]{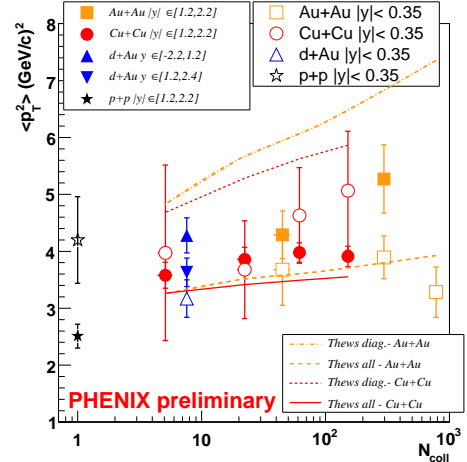}
\end{center}
\end{minipage}
\vspace{-0.4cm}
\caption{\label{y_pt_distribution}{Left: J/$\psi$ rapidity distribution for p+p, Cu+Cu (red circles) and Au+Au (yellow squares) collisions for different centrality bins. Right: $\langle p_t^2 \rangle$ for p+p (black stars), d+Au (blue triangles), Cu+Cu (red circles) and Au+Au (yellow squares) as a function of centrality. Filled markers correspond to the forward rapidity measurements while open markers to the mid-rapidity measurement. Predictions~{\protect\cite{thews_01}}: with full recombination scenario for Cu+Cu collisions (red solid line) and for Au+Au (yellow dashed line) and no recombination for Cu+Cu (red dotted line) and for Au+Au (yellow dot-dashed line).}}
\vspace{-0.4cm}
\end{figure}

Models without charm quark recombination generally assume that J/$\psi$ production takes place at the early stage of the collisions. In this frame, subsequent interactions in the medium (whatever its nature) can only lead to a decrease of the final J/$\psi$ yield. At RHIC energy, multiple $c\bar{c}$ pairs are produced in central collisions~\cite{phenix_charm} and quark mobility in a deconfined medium allows uncorrelated charm quarks to recombine and to form a J/$\psi$. Such a recombination could at least partially compensate J/$\psi$ suppression. However, recombination scenarios imply modifications of the rapidity and transverse momentum distributions. Recombined J/$\psi$ production is proportional to the square of the $c\bar{c}$ pair yields and one expects that recombined production will be peaked in the region where the $c\bar{c}$ pairs are mainly produced: at mid rapidity and at low $p_t$ according to pQCD predictions. As a consequence, the rapidity distribution and the transverse momentum distribution are expected to be narrower in central collisions. The left figure~\ref{y_pt_distribution} presents the rapidity distribution for p+p, Cu+Cu and Au+Au collisions for different centrality bins: within the error bars, there is no evidence of a modification of the rapidity distribution. The right figure~\ref{y_pt_distribution} presents the $\langle p^2_t \rangle$ as a function of centrality. The results are in between predictions with full and no recombination, suggesting that the measurements consist of a mixture of recombined and non recombined J/$\psi$.\\

\begin{figure}[htbp!]
\vspace{-0.3cm}
\begin{minipage}{7.8cm}
\begin{center}\includegraphics[width=6.5cm,keepaspectratio]{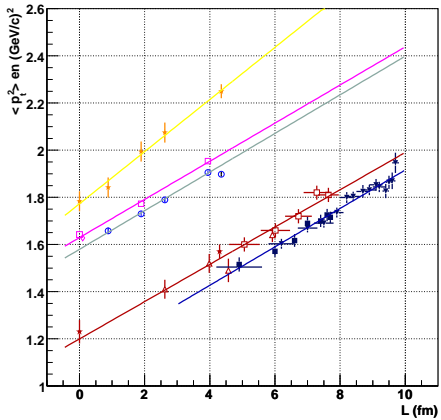}
\end{center}
\end{minipage}
\hspace{0.1cm}
\begin{minipage}{7.7cm}
\begin{center}
\includegraphics[width=6.5cm,keepaspectratio]{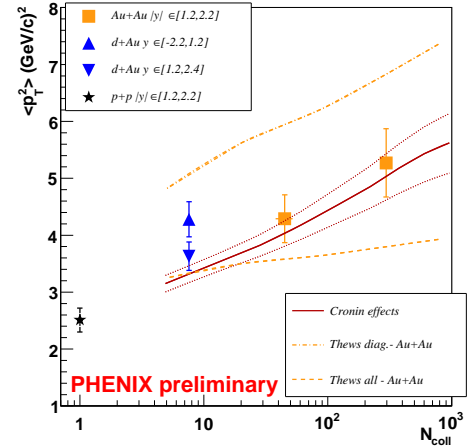}
\end{center}
\end{minipage}
\vspace{-0.4cm}
\caption{\label{sps_rhic_pt_square}{Left: $\langle p_t^2 \rangle$ as a function of L (fm). Solid blue markers correspond to Pb+Pb collisions (at 158 GeV fixed target) at NA50~{\protect\cite{na50_pt}}, and In+In collisions at NA60~{\protect\cite{pillot}}, red markers correspond to p+p collisions at NA3~{\protect\cite{badier}} and p+Cu, p+U, O+Cu, O+U and S+U at NA38~{\protect\cite{na38}} (at 200 GeV fixed target), open blue markers correspond to p+Be, p+Al, p+Cu, p+W and p+Pb collisions in NA50~{\protect\cite{na50_pt}} (at 400 GeV fixed target), pink markers correspond to p+p et p+d collisions in NA51 and p+Al et p+W collisions in NA50 (at 400~GeV fixed target), orange markers are measurements from E866/789/771~{\protect\cite{leitch}}at 800~GeV (fixed target). Right: $\langle p_t^2 \rangle$ as a function of centrality. Red solid line (error band in dotted lines) corresponds to the Cronin effect expectation using Eq.~\ref{eqn:cronin} and results from~{\protect\cite{ppg038}}.}}
\vspace{-0.4cm}
\end{figure}

We present here an alternative description of the observed $p_t$ broadening using the Cronin effect. The broadening is described as a consequence of gluon-ncucleon scattering in the initial state. In principle, this effect leads to:
\begin{eqnarray}
\langle p_t^2 \rangle_{AA} = \langle p_t^2 \rangle_{pp} + \rho \sigma \delta( \langle p_t^2 \rangle) \times L_{AA}
\label{eqn:cronin}
\end{eqnarray}
where $\langle p_t^2 \rangle_{pp}$ is the $\langle p_t^2 \rangle$ in p+p collision, $\rho$ is the  nuclear density, $\sigma$ is the elastic gluon-nucleon scattering cross section, $\delta( \langle p_t^2 \rangle)$ is the kick given by each scattering and $L_{AA}$ is the average thickness of nuclear matter of the centrality bin considered. Eq~\ref{eqn:cronin} gives a reasonable description $\langle p_t^2 \rangle$ measurements at lower energy experiments as shown by left figure~\ref{sps_rhic_pt_square} as a function of L. The right figure~\ref{sps_rhic_pt_square} presents the extrapolation (from p+p and d+Au) of the Cronin effect, showing that the $p_t$ broadening in Au+Au measured at forward rapidity 
 can be explained with this effect alone. This suggests this anomalous J/$\psi$ suppression observed in the most central collisions does not affect the $p_t$ distribution. However, due to error bars, it is difficult to distinguish between different models and to make a strong statement about the $p_t$ broadening.


\section*{References}
\begin{thebibliography}{99}
\bibitem{psi_result_na50} NA50 collaboration, Eur. Phys. J.C.39, 335.
\bibitem{comovers} A. Capella, D. Sousa, nucl-th/0303055 (2003).
\bibitem{hugo} H. Pereira, nucl-ex/0510051.
\bibitem{ppg038} PHENIX collaboration, \PRL{96, 012304}.
\bibitem{vogt} R. Vogt, nucl-th/0507027 (2005) + private communications.
\bibitem{vogt_dAu} R. Vogt, \PRC{71, 054902}
\bibitem{na51} NA50 collaboration, \PLB{553, 167}.
\bibitem{phenix_charm} PHENIX collaboration, \PRL{94, 082301}.
\bibitem{thews_01} Robert L. Thews, \PRC{73 014904} + private communications. 
\bibitem{na50_pt} MC. Abreu et al, Phys. Lett. B499 (2001) 85-86.
\bibitem{pillot} P. Pillot PhD. thesis (2005) Institut de Physique Nucl\'eaire de Lyon (IPNL).
\bibitem{badier} Badier et al, Z Phys. C20 (1984) 101.
\bibitem{na38} C. Baglin et al, \PLB{262, 362} and MC. Abreu et al, \PLB{423, 207}.
\bibitem{leitch} M. Leitch, Eur Phys. J. A19 s01 (2004) 129 and \PRL{84, 3256}.
\end{thebibliography}
\end{document}